\documentclass[12pt]{article}
\usepackage{latexsym,graphicx,amssymb}

\def\bseq{\begin{subequation}}  
\def\eseq{\end{subequation}}
\def\bsea{\begin{subeqnarray}}  
\def\esea{\end{subeqnarray}}

\newcommand{\bbox}{\lower.2ex\hbox{$\Box$}}

\newcommand{\beq}{\begin{equation}}
\newcommand{\eeq}{\end{equation}}
\newcommand{\bea}{\begin{eqnarray}}
\newcommand{\eea}{\end{eqnarray}}
\newcommand{\ena}{\end{eqnarray}}
\newcommand{\ba}{\begin{array}}
\newcommand{\ea}{\end{array}}
\newcommand{\ben}{\begin{enumerate}}
\newcommand{\een}{\end{enumerate}}
\newcommand{\bde}{\begin{description}}
\newcommand{\ede}{\end{description}}

\renewcommand{\a}{\alpha}
\renewcommand{\b}{\beta}

\renewcommand{\d}{\delta}

\newcommand{\pa}{\partial}
\newcommand{\g}{\gamma}
\newcommand{\G}{\Gamma}

\newcommand{\e}{\epsilon}

\renewcommand{\l}{\lambda}

\newcommand{\p}{\pi}

\renewcommand{\r}{\right}
\renewcommand{\l}{\left}

\newcommand{\Phib}{\bar{\Phi}}

\newcommand{\ad}{{\dot{\alpha}}}
\newcommand{\bd}{{\dot{\beta}}}

\begin{document}
\begin{titlepage}
\begin{flushright}
IFUM-799-FT\\
July 2004
\end{flushright}
\vspace{2cm}

\noindent{\Large \bf Two-point functions for ${\cal{N}}=4$ Konishi-like operators}
\vspace{.5cm}
{\bf \hrule width 16.cm}
\vspace {1cm}

\noindent{\large \bf Stefano Maghini, Alberto Santambrogio and Daniela Zanon}

\vskip 2mm

{ \small \noindent Dipartimento di Fisica dell'Universit\`a di Milano and

\noindent INFN, Sezione di Milano, Via Celoria 16, 20133 Milano, Italy}
\vfill
\begin{center}
{\bf Abstract}
\end{center}
{\small We compute the two-point function of Konishi-like
operators up to one-loop order, in ${\cal{N}}=4$ supersymmetric
Yang-Mills theory. We work perturbatively in ${\cal N}=1$
superspace.  We find the expression expected on the basis of
superconformal invariance and determine the normalization of the
correlator and the anomalous dimension of the operators to order
$g^2$ in the coupling constant. }
\vspace{2mm} \vfill \hrule width
6.cm
\begin{flushleft}
e-mail: stefano.maghini@mi.infn.it\\
e-mail: alberto.santambrogio@mi.infn.it\\
e-mail: daniela.zanon@mi.infn.it
\end{flushleft}
\end{titlepage}

Recently the ${\cal N}=4$ supersymmetric Yang-Mills theory has
attracted a lot of attention primarily in connection with the
AdS/CFT correspondence \cite{adscft}. Actually this theory is
interesting by itself, being the prototype of a superconformal
field theory in 4 dimensions \cite{SW}, and there is the hope that
the problem of solving such a theory might be an attackable one.

In the past ${\cal N}=1$ superspace techniques have been used in
order to capture relevant insights on the structure of this theory
\cite{GKP,PSZ,BKRS,PS,APPSS,SZ}, e.g. the computation of
correlation functions or the evaluation of the anomalous
dimensions of conformal operators as functions of the coupling
constant in perturbation theory. Of course the aim would be to
obtain exact results, i.e. one would like to determine the full
dependence on the parameters that appear in the theory. Despite
the fact that one deals with a perturbative expansion there is
growing evidence that in certain sectors of the theory one might
be able to rearrange the perturbative calculation and succeed in
summing the series (see, e.g., \cite{RyzTse} and references
therein).

Our project is to see how far we can get combining these standard
superspace Feynman rules with general knowledge about the
conformal invariance of the theory. In this paper, as a first
simple application, we want to test these techniques in the
calculation of one-loop two-point functions for a class of
non-protected operators which are a generalization of  the well
studied Konishi operator. The general form of the two-point
function is fixed by conformal invariance. We will show how to
obtain the logarithmic part and also the part corresponding to the
normalization of the two-point function, which in general is also
renormalized, up to order $g^2$ in the coupling.

\vspace{0.8cm}

The ${\cal N}=4$ supersymmetric Yang-Mills classical action
written in terms of ${\cal N}=1$ superfields (we use the notations
and conventions adopted in \cite{superspace}) is given by
\bea\label{N4SYMaction}
S & = & {\rm Tr} \l( \int d^4x~d^4 \theta \, e^{-gV} \bar{\Phi}_i \, e^{\,gV} \Phi^i + \frac{1}{2 g^2} \int d^4x~d^2 \theta ~ W^\a W_\a + \frac{1}{2 g^2} \int d^4x~d^2 \bar{\theta} ~ \bar{W}^\ad \bar{W}_\ad \r. \nonumber \\
&& \left. ~~~~~ + \frac{g}{3!} \int d^4x~d^2 \theta ~ i \e_{ijk} \, \Phi^i \l[\Phi^j,\Phi^k\r] + \frac{g}{3!} \int d^4x~d^2 \bar{\theta} ~ i \e^{ijk} \, \bar{\Phi}_i \l[\bar{\Phi}_j,\bar{\Phi}_k\r] \r)
\eea
where the $\Phi^i$ with $i=1,2,3$ are
three chiral superfields, and the $W^\a= i \bar{D}^2 \l(e^{-gV}D^\a \, e^{\,gV}\r)$
are the gauge superfield strengths. All the fields are
Lie-algebra valued, e.g. $\Phi^i=\Phi^i_a T^a$, in the adjoint
representation of the gauge group.

We consider the following operators, for $k \geq 1$
\beq\label{konishi}
{\cal{O}} = {\rm Tr} \l(\sum_{i=1}^3 e^{-gV} \bar{\Phi}_i \, e^{\,gV} \Phi^{(i} \Phi^{j_1} \cdots \Phi^{j_{k-1})}\r)
\qquad
\bar{\cal{O}} = {\rm Tr} \l(\sum_{i=1}^3 e^{\,gV} \Phi^i \, e^{-gV} \bar{\Phi}_{(i} \bar{\Phi}_{j^\prime_1} \cdots \bar{\Phi}_{j^\prime_{k-1})}\r) \nonumber \\
\eeq
We want to compute the correlators
\beq
{\cal{K}}\l(z,z^\prime\r) \equiv \, <{\cal{O}}\l(x,\theta\r) \bar{\cal{O}}\l(x^\prime,\theta^\prime\r)>
\label{2point} \eeq to order $g^2$ in the coupling constant.

${\cal N}=4$ super Yang-Mills is a superconformal field theory and
we know that superconformal invariance fixes the form of the two-
and three-point functions of primary operators, up to a
normalization constant, in terms of their dimensions and chiral
weights \cite{PO}. For the two-point function, a general
expression was given in terms of ${\cal N}=1$ superfields in
\cite{SZ}
\beq
<{\cal{Q}}\l(z\r) \bar{{\cal{Q}}}\l(z^\prime\r)> \,
= f_{{\cal{Q}}} \left\{\frac{1}{2}D^\a \bar D^2 D_\a +
\frac{1}{4}\frac{w}{\Delta} \l[ D^\a,\bar D^{\ad} \r] i
\pa_{\a\ad} + \frac{1}{4}\frac{\Delta^2 + w^2 -2\Delta}{\Delta
\l(\Delta - 1\r)}\Box\right\}
\frac{\delta^4\l(\theta-\theta^\prime\r)}{{\l(x-x^\prime\r)}^{2\Delta}}
\label{formula}
\eeq

\vspace{.3cm} \noindent where $\Delta$ is the total dimension of
the operator ${\cal Q}$, $w$ is the chiral weight and
$f_{{\cal{Q}}}$ is the arbitrary normalization constant. The total
dimension is the sum of the classical plus the anomalous
dimension, $\Delta=\Delta_0 + \gamma$, while the chiral weight is
not renormalized because of ${\cal N}=4$ supersymmetry.

Our operators in (\ref{konishi}) have $\Delta_0 = k+1$ and $w=k-1$. At one-loop
order we write
\beq
\Delta=k+1+\gamma \qquad w=k-1 \qquad f_{\cal O}=A+Bg^2
\eeq
with $\gamma=O\l(g^2\r)$. By using these values in (\ref{formula}) and
expanding up to order $g^2$ a straightforward calculation gives
\bea\label{formula1loop}
<{\cal{O}}\l(z\r) \bar{{\cal{O}}}\l(z^\prime\r)> & = & A \left\{ \bar D^2 D^2 + \frac{1}{k+1} {\bar{D}}^{\ad} D^\a i\pa_{\a\ad} \right\} \frac{\delta^{(4)}\l(\theta-\theta^\prime\r)}{{\l(x-x^\prime\r)}^{2\l(k+1\r)}} \nonumber \\
&& - ~ A \gamma \left\{ \bar D^2 D^2 + \frac{1}{k+1} \bar D^{\ad} D^\a i\pa_{\a\ad} \right\} \frac{\delta^{(4)}\l(\theta-\theta^\prime\r)}{{\l(x-x^\prime\r)}^{2\l(k+1\r)}} \log{\l(x-x^\prime\r)}^2 \nonumber \\
&& + ~ B g^2 \left\{ \bar D^2 D^2 + \frac{1}{k+1} \bar D^{\ad} D^\a i\pa_{\a\ad} \right\} \frac{\delta^{(4)}\l(\theta-\theta^\prime\r)}{{\l(x-x^\prime\r)}^{2\l(k+1\r)}} \nonumber \\
&& + ~ A \gamma \frac{1}{2\l(k+1\r)} \left\{ \frac{k-1}{k+1} \bar D^{\ad} D^\a i\pa_{\a\ad} + \frac{1}{k} \Box \right\} \frac{\delta^{(4)}\l(\theta-\theta^\prime\r)}{{\l(x-x^\prime\r)}^{2\l(k+1\r)}}
\eea
The first row of this formula gives the
classical part of the two-point function, while the rest
corresponds to the one-loop contribution.

In this paper we will reproduce the expected expression
(\ref{formula1loop}) by explicitly computing the two-point
function up to one-loop. To this order we will determine the
normalization of the two-point function and the anomalous
dimension of the operators.\footnote{Here and in the rest of the
paper, with ``anomalous dimension" we simply mean the coefficient
of the log term in the two-point function, which in general is not
an eigenvalue of the dilatation operator due to the mixing among
our operators and many others with the same classical dimension.
Solving this mixing problem would be an unnecessary complication
for our purposes of reproducing formula (\ref{formula1loop}), and
moreover this was already done in \cite{BERS,B}}.

The calculation is most easily performed using ${\cal{N}}=1$
superspace techniques: we introduce sources in the action
(\ref{N4SYMaction}) as
\beq
\int d^4x~d^4 \theta \l({\cal{O}}J+\bar{\cal{O}}\bar{J}\r)
\eeq
and define the generating functional in Euclidean space
\beq
W\l[J,\bar{J} \, \r]=\int {\cal D}\Phi~{\cal D}\Phib~{\cal D}V~e^{S\,[J,\bar{J}\,]}
\label{genfunc}
\eeq
so that
the two-point function is given by
\beq
<{\cal O}\l(z\r) \bar{{\cal O}}\l(z^\prime\r)> \, = \left. \frac{\d^2 W}{\d J\l(z\r)\d\bar{J}\l(z^\prime\r)}\right|_{J=\bar{J}=0}
\label{defcorr}
\eeq
We use perturbation theory to evaluate the contributions to $W\l[J,\bar{J}\,\r]$
which are
quadratic in the sources, i.e.
\beq
W\l[J,\bar{J} \, \r] \rightarrow \int d^4x ~ d^4x^\prime ~ d^4 \theta~ J\l(x,\theta,\bar{\theta}\r) \frac{F\l(g^2\r)}{{\l(x-x^\prime\r)}^{2\l(k+1\r)}}\bar{J}\l(x^\prime,\theta,\bar{\theta}\r)
\label{twopoint}
\eeq
As mentioned above the $x-$dependence
of the result is fixed by the conformal invariance of the theory,
and $F\l(g^2\r)$ is the function to be determined.

In order to obtain
the result in (\ref{twopoint}) one has to consider all the two-point
diagrams from $W\l[J,\bar{J}\,\r]$ with $J$ and $\bar{J}$ on the external
legs.
The rules are standard: since
at one-loop we will have to deal with divergent integrals, we find
it convenient to use dimensional regularization in $d=4-2\e$
 $x-$space within the G-scheme \cite{russians}. We need not
worry about supersymmetric dimensional reduction since, as we will
see, the potentially dangerous diagrams are finite. Thus we use as superfield propagators
\bea
<\Phi^{ia}\l(x,\theta\r) \bar{\Phi}^b_{~j}\l(x^\prime,\theta^\prime\r)> & = & \d^i_{~j} \, \d^{ab} \, \frac{\G\l(1-\e\r)}{4\p^{2-\e}}~\frac{\d^{(4)}\l(\theta-\theta^\prime\r)}{{\l(x-x^\prime\r)}^{2\l(1-\e\r)}} \nonumber \\
\nonumber \\
<V^a\l(x,\theta\r) V^b\l(x^\prime,\theta^\prime\r)> & = & - \, \d^{ab} \, \frac{\G\l(1-\e\r)}{4\p^{2-\e}}~\frac{\d^{(4)}\l(\theta-\theta^\prime\r)}{{\l(x-x^\prime\r)}^{2\l(1-\e\r)}}
\label{propagators}
\eea
while the relevant vertices are read directly from the action (\ref{N4SYMaction}) and the expressions
of the operators (\ref{konishi}).

At tree level we obtain
\bea
{\cal{K}}_0\l(0,z\r) & = & \frac{1}{{\l(2\p\r)}^{2\l(k+1\r)}} \l(k+2\r) \l(k-1\r)! ~ \d^{j_1}_{(j^\prime_1} \cdots \d^{j_{k-1}}_{j^\prime_{k-1})} \nonumber \\
&&{\rm Tr} \l(T_{(a_1} \cdots T_{a_{k+1})}\r) {\rm Tr} \l(T_{(a_1} \cdots T_{a_{k+1})}\r) \frac{\d^{(4)}\l(\theta\r)}{x^{2\l(k+1\r)}}
\label{treelevel}
\eea

\begin{figure}[h!tb]
\begin{center}
\includegraphics[scale=.95]{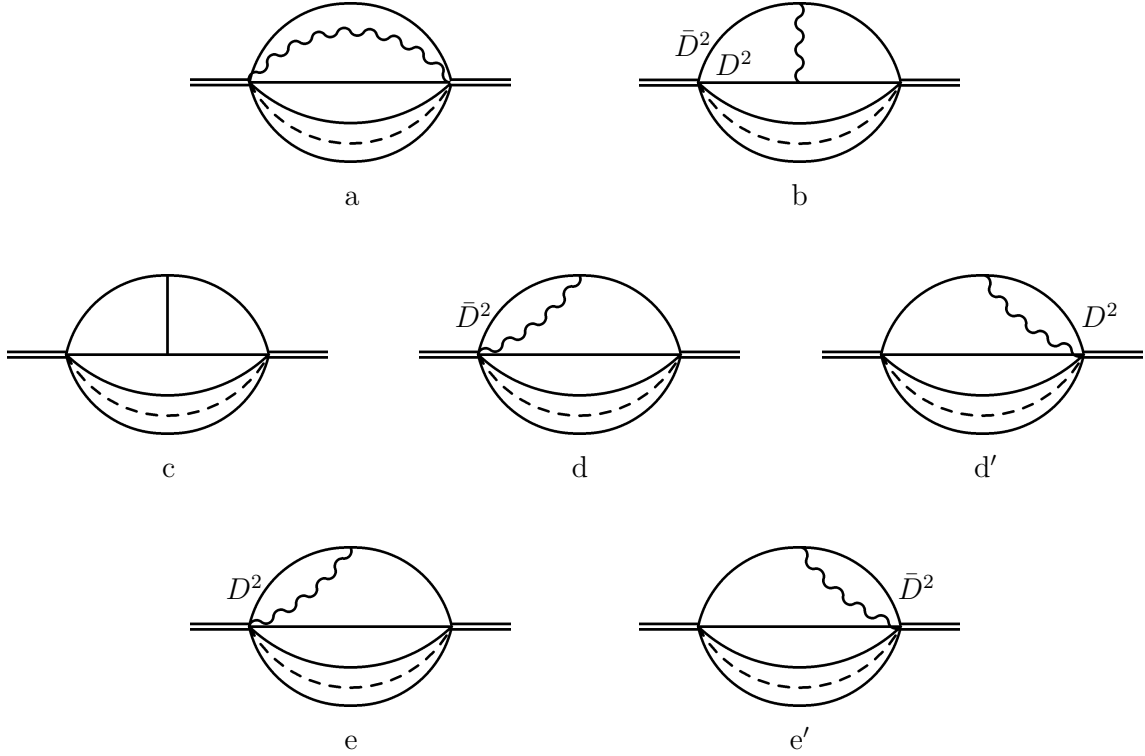}
\caption{Diagrams whose $D-$algebra gives relevant contributions.}\label{Relevantcontributions}
\end{center}
\vspace{.8cm}
\end{figure}

\begin{figure}[h!tb]
\begin{center}
\includegraphics[scale=.95]{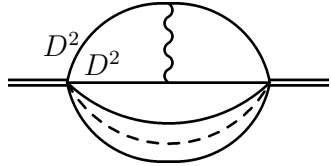}
\caption{Diagram not vanishing, but not relevant (proportional to contact terms).}\label{Not_vanishing_but_not_relevant}
\end{center}
\end{figure}

\vspace{0.8cm}
At one-loop the relevant supergraphs are shown in Fig. \ref{Relevantcontributions}.
We have not drawn diagrams which give a vanishing contribution as a result of $D-$algebra
or color symmetry factors. In addition
we do not consider diagrams like the one shown in Fig. \ref{Not_vanishing_but_not_relevant} which give rise to contact terms. Notice
that we have not included self-energy insertions since for the ${\cal N}=4$ Yang-Mills theory there
are no one-loop propagator corrections.

\vspace{1cm}

\newpage

First we perform the superspace $D-$algebra and reduce the result to a multi-loop integral, then we evaluate
all factors coming from combinatorics and color structures of a given diagram.

\begin{figure}[h!tb]
\begin{center}
\includegraphics[scale=.95]{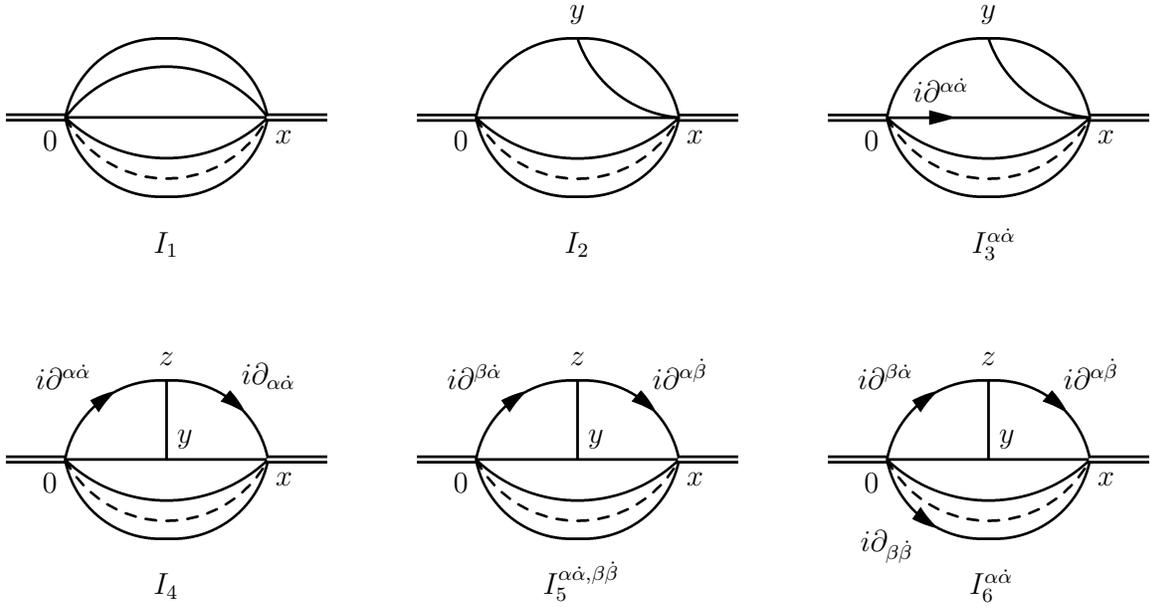}
\caption{Bosonic contributions obtained after $D-$algebra.}\label{Integrals}
\end{center}
\end{figure}

The $D-$algebra leads to the following results:
\bea\label{Dalgebra}
(a) & \rightarrow & J \l( 0 , \theta , \bar{\theta} \r) \l[ - I_1\l(x\r) \r] \bar{J} \l( x , \theta , \bar{\theta} \r) \nonumber\\
\nonumber \\
(b) & \rightarrow & J \l( 0 , \theta , \bar{\theta} \r) \l[ 2 I_2\l(x\r) \bar{D}^2 D^2 + 2 I_3^{\a\ad}\l(x\r) \bar{D}_\ad D_\a + I_4\l(x\r) \bar{D}^2 D^2 \r. \nonumber \\
&& \l. \hspace{1.9cm} - I_5^{\a\ad,\b\bd}\l(x\r) i \pa_{\b\bd}\bar{D}_\ad D_\a + \l(k-1\r) I_6^{\a\ad}\l(x\r) \bar{D}_\ad D_\a \r] \bar{J} \l( x , \theta , \bar{\theta} \r) \nonumber \\
\nonumber \\
(c) & \rightarrow & J \l( 0 , \theta , \bar{\theta} \r) \l[ I_1\l(x\r) - 2 I_2\l(x\r) \bar{D}^2 D^2 - 2 I_2\l(x\r) i\pa_{\a\ad}\bar{D}^\ad D^\a \r. \nonumber \\
&& \hspace{1.9cm} + 2 \, k \, I_3^{\a\ad}\l(x\r) \bar{D}_\ad D_\a + I_4\l(x\r) \bar{D}^2 D^2 - I_5^{\a\ad,\b\bd}\l(x\r) i\pa_{\b\bd}\bar{D}_\ad D_\a \nonumber \\
&& \l. \hspace{1.9cm} + \l( k-1 \r) I_6^{\a\ad}\l(x\r) \bar{D}_\ad D_\a \r] \bar{J} \l( x , \theta , \bar{\theta} \r) \nonumber \\
\nonumber \\
(d) + (d^\prime) & \rightarrow & J \l( 0 , \theta , \bar{\theta} \r) \l[ -2 I_2\l(x\r) \bar{D}^2 D^2 - 2 I_3^{\a\ad}\l(x\r) \bar{D}_\ad D_\a \r] \bar{J} \l( x , \theta , \bar{\theta} \r) \nonumber \\
\nonumber \\
(e) + (e^\prime) & \rightarrow & J \l( 0 , \theta , \bar{\theta} \r) \l[ 2 I_1\l(x\r) - 2 I_2\l(x\r) \bar{D}^2 D^2 - 2 I_2\l(x\r) i\pa_{\a\ad} \bar{D}^\ad D^\a \r. \nonumber \\
&& \l. \hspace{1.9cm} + 2 \, k \, I_3^{\a\ad}\l(x\r) \bar{D}_\ad D_\a \r] \bar{J} \l( x , \theta , \bar{\theta} \r)
\eea
where $I_1,\ldots,I_6^{\a\ad}$ correspond to the bosonic graphs in Fig. \ref{Integrals}.

\vspace{.2cm}

They can be written explicitly using dimensional regularization in $d=4-2\e$ $x-$space:
\bea\label{oneloop}
I_1 & = & {\l( \frac{\G\l(1-\e\r)}{4\p^{2-\e}}\r)}^{k+2} \frac{1}{x^{2\l(k+2\r)\l(1-\e\r)}} \nonumber \\
\nonumber \\
\nonumber \\
I_2 & = & {\l( \frac{\G\l(1-\e\r)}{4\p^{2-\e}}\r)}^{k+3} \frac{1}{x^{2k\l(1-\e\r)}} \int d^{4-2\e}y ~ \frac{1}{y^{2\l(1-\e\r)}{\l(y-x\r)}^{4\l(1-\e\r)}} \nonumber \\
\nonumber \\
\nonumber \\
I_3^{\a\ad} & = & {\l( \frac{\G\l(1-\e\r)}{4\p^{2-\e}}\r)}^{k+3} \frac{1}{x^{2\l(k-1\r)\l(1-\e\r)}} \l(-i \pa^{\,\a\ad}\frac{1}{x^{2\l(1-\e\r)}}\r) \int d^{4-2\e}y ~ \frac{1}{y^{2\l(1-\e\r)}{\l(y-x\r)}^{4\l(1-\e\r)}} \nonumber \\
\nonumber \\
\nonumber \\
I_4 & = & {\l( \frac{\G\l(1-\e\r)}{4\p^{2-\e}}\r)}^{k+4} \frac{1}{x^{2\l(k-1\r)\l(1-\e\r)}} \int d^{4-2\e}y ~ d^{4-2\e}z ~ \l( -i \pa^{\,\a\ad} \frac{1}{z^{2\l(1-\e\r)}}\r) \nonumber \\
&& \hspace{5.1cm} \l( i \pa_{\,\a\ad} \frac{1}{{\l(z-x\r)}^{2\l(1-\e\r)}}\r) \frac{1}{{\l[ y \l(y-z\r)\l(y-x\r)\r]}^{2\l(1-\e\r)}} \nonumber \\
\nonumber \\
\nonumber \\
I_5^{\a\ad,\b\bd} & = & {\l( \frac{\G\l(1-\e\r)}{4\p^{2-\e}}\r)}^{k+4} \frac{1}{x^{2\l(k-1\r)\l(1-\e\r)}} \int d^{4-2\e}y ~ d^{4-2\e}z ~ \l( -i \pa^{\,\b\ad} \frac{1}{z^{2\l(1-\e\r)}}\r) \nonumber \\
&& \hspace{5.1cm} \l( i \pa^{\,\a\bd}\frac{1}{{\l(z-x\r)}^{2\l(1-\e\r)}}\r) \frac{1}{{\l[y\l(y-z\r)\l(y-x\r)\r]}^{2\l(1-\e\r)}} \nonumber \\
\nonumber \\
\nonumber \\
I_6^{\a\ad} & = & {\l( \frac{\G\l(1-\e\r)}{4\p^{2-\e}}\r)}^{k+4} \frac{1}{x^{2\l(k-2\r)\l(1-\e\r)}} \l( - i\pa_{\,\b\bd} \, \frac{1}{x^{2\l(1-\e\r)}}\r) \int d^{4-2\e}y ~ d^{4-2\e}z ~ \l( -i \pa^{\,\b\ad} \frac{1}{z^{2\l(1-\e\r)}}\r) \nonumber \\
&& \hspace{5.1cm} \l( i \pa^{\,\a\bd}\frac{1}{{\l(z-x\r)}^{2\l(1-\e\r)}}\r) \frac{1}{{\l[y\l(y-z\r)\l(y-x\r)\r]}^{2\l(1-\e\r)}} \nonumber \\
\eea
With our conventions we have
\bea
x^2 & \equiv & 2x^{\a\ad}x_{\a\ad} \nonumber \\
x^{\a\ad} x^\b_\ad & = & \frac{1}{4} C^{\b\a} x^2 \nonumber \\
i \pa^{\,\a\ad} \frac{1}{x^{2n}} & = & -4n \frac{ix^{\a\ad}}{x^{2\l(n+1\r)}}
\label{conventions}
\eea
The $I_4$, $I_5^{\a\ad,\b\bd}$ and $I_6^{\a\ad}$ contributions are finite.
They can be computed (see for example \cite{GZ}) evaluating the following integral in the
$\e\rightarrow 0$ limit
\vspace{.4cm}
\bea
{\l( \frac{\G\l(1-\e\r)}{4\p^{2-\e}}\r)}^5 \int d^{4-2\e}y ~ d^{4-2\e}z ~ \l( i \pa^{\,\a\ad} \frac{1}{z^{2\l(1-\e\r)}}\r) \l( i \pa^{\,\b\bd}\frac{1}{{\l(z-x\r)}^{2\l(1-\e\r)}}\r) \frac{1}{{\l[y\l(y-z\r)\l(y-x\r)\r]}^{2\l(1-\e\r)}} \nonumber
\eea
\beq
\longrightarrow - \frac{1}{12{\l(2\p\r)}^6} \frac{1}{x^4} \l(C^{\a\b} C^{\ad\bd} + 4 \frac{x^{\a\ad} x^{\b\bd}}{x^2} \r)
\eeq

\vspace{.5cm} \noindent We notice that these are the terms that
one would need to worry about maintaining supersymmetry via
regularization: indeed supersymmetric regularization would tell us
to use the dimensional reduction rule $C^{\a\b}C_{\a\b}=2$ as
opposed to the dimensional regularization rule
$C^{\a\b}C_{\a\b}=2-\e$. Of course, being these integrals finite,
any rule is a good rule. The $I_2$ and $I_3^{\a\ad}$ contributions
are divergent but do not contain potentially dangerous
contractions of indices. Thus one computes the integrals and
subtracts subdivergences in a standard manner. With an overall
common factor \beq \frac{1}{{\l(2\p\r)}^{2\l(k+2\r)}}
\label{factor} \eeq we obtain the following finite results, up to
integrations by parts \bea
I_1 & = & \frac{1}{x^{2\l(k+2\r)}} \, = \, \frac{1}{4k\l(k+1\r)} \, \Box \, \frac{1}{x^{2\l(k+1\r)}} \nonumber \\
\nonumber \\
\nonumber \\
I_2 & = & - \frac{G\l(x\r)}{4} \frac{1}{x^{2\l(k+1\r)}} \nonumber \\
\nonumber \\
\nonumber \\
I_3^{\a\ad} \bar{D}_\ad D_\a & = & - \frac{ix^{\a\ad}}{x^{2\l(k+2\r)}} G\l(x\r) \bar{D}_\ad D_\a \, = \, \frac{1}{4\l(k+1\r)} \l( i \pa^{\a\ad} \frac{1}{x^{2\l(k+1\r)}}\r) G\l(x\r) \bar{D}_\ad D_\a \nonumber \\
& = & - \frac{1}{4\l(k+1\r)} \frac{1}{x^{2\l(k+1\r)}} G\l(x\r) i \pa^{\a\ad} \bar{D}_\ad D_\a + \frac{1}{4{\l(k+1\r)}^2} \frac{1}{x^{2\l(k+1\r)}} i \pa^{\a\ad} \bar{D}_\ad D_\a \nonumber \\
\nonumber \\
\nonumber \\
I_4 & = & \frac{1}{2} \frac{1}{x^{2\l(k+1\r)}} \nonumber \\
\nonumber \\
\nonumber \\
I_5^{\a\ad,\b\bd} i \pa_{\b\bd} \bar{D}_\ad D_\a & = & -\frac{1}{12} \frac{1}{x^{2\l(k+1\r)}} i \pa^{\a\ad} \bar{D}_\ad D_\a - \frac{1}{3} i \pa^{\b\bd} \l( \frac{x_{\b\ad}x_{\a\bd}}{x^{2\l(k+2\r)}} \r) \bar{D}^\ad D^\a \nonumber \\
& = & - \frac{1}{4\l(k+1\r)} \frac{1}{x^{2\l(k+1\r)}} i \pa^{\a\ad} \bar{D}_\ad D_\a \nonumber \\
\nonumber \\
\nonumber \\
I_6^{\a\ad} \bar{D}_\ad D_\a & = & -\frac{1}{12} \frac{1}{x^{2k}} \l( i \pa_{\b\bd}\frac{1}{x^2} \r) \l( C^{\a\b} C^{\bd\ad} + 4 \frac{x^{\b\ad} x^{\a\bd}}{x^2} \r) \bar{D}_\ad D_\a \nonumber \\
& = & -\frac{1}{12\l(k+1\r)} \frac{1}{x^{2\l(k+1\r)}} i \pa^{\a\ad} \bar{D}_\ad D_\a + \frac{1}{12\l(k+1\r)} \frac{1}{x^{2\l(k+1\r)}} i \pa^{\a\ad} \bar{D}_\ad D_\a \, = \, 0 \nonumber \\
\eea
where we have defined $G\l(x\r)=\a - \log{x^2}$ with $\a$ a scheme dependent constant.

\vspace{0.8cm}
Having performed the $D-$algebra and computed the resulting bosonic integrals, in order to
obtain the complete $g^2$ contribution to the  two-point correlator we need insert the color
and symmetry factors for each diagram in Fig. \ref{Relevantcontributions}.

We notice that a non trivial check of the $D-$algebra can be done
specializing the result for chiral primary CPO-type operators in
which case we know that the total answer must give a vanishing
result. We consider such an operator, e.g. \beq\label{CPO}
{\cal{C}} \equiv \bar{\Phi}_1 \Phi^{(2} \Phi^3 \cdots \Phi^{3)}
\eeq where the total number of fields is again $k+1$. Then if we
compute the correlator \beq <{\cal{C}}\l(0\r)
\bar{\cal{C}}\l(z\r)> \label{2pointCPO} \eeq to order $g^2$, we
will have the same type of diagrams as in Fig.
\ref{Relevantcontributions}  with the result of the $D-$algebra
given in (\ref{Dalgebra}). The color and symmetry factors
$f_{(i)}$ for the CPO operators in (\ref{CPO}) are easily computed
leading to \beq
f_{(a)}=f_{(b)}=-f_{(c)}=f_{(d)}=f_{(d^\prime)}=f_{(e)}=f_{(e^\prime)}
\label{factorCPO} \eeq It is immediate to check that using
(\ref{factorCPO}) in (\ref{Dalgebra}) the contributions from the
various diagrams exactly cancel.

\vspace{0.8cm}

Now we turn to the computation of the color and symmetry factors
$F_{(i)}$ for the Konishi-type correlator. We find \beq
F_{(a)}=F_{(b)}=\frac{k}{2}F_{(c)}=F_{(d)}=F_{(d^\prime)}=F_{(e)}=F_{(e^\prime)}=2g^2N
F_{tree} \label{factorK} \eeq where we have assumed gauge group
$SU(N)$, and \beq F_{tree}=\l(k+2\r)\l(k-1\r)! ~
\d^{j_1}_{(j^\prime_1} \cdots \d^{j_{k-1}}_{j^\prime_{k-1})} ~
{\rm Tr} \l(T_{(a_1} \cdots T_{a_{k+1})}\r) {\rm Tr} \l(T_{(a_1}
\cdots T_{a_{k+1})}\r) \label{factortreelevel} \eeq is the color
and symmetry factor which appears in the tree-level computation in
(\ref{treelevel}). Now we can assemble the various contributions
for the Konishi-type two-point function: with an overall factor
\beq \frac{1}{{\l(2\p\r)}^{2\l(k+2\r)}} 2 g^2 N F_{tree}
\label{factorKfinal} \eeq we have \bea
&& \l( 1 + \frac{2}{k}\r) J \l( 0,\theta,\bar \theta \r) \l[ I_1\l(x\r) - 2 I_2\l(x\r) \l( \bar{D}^2 D^2 + i \pa_{a\ad} \bar{D}^\ad D^\a \r) + 2 \, k \, I_3^{\a\ad}\l(x\r) \bar{D}_\ad D_\a \r. \nonumber \\
&& \hspace{2cm} + \l. + I_4\l(x\r) \bar{D}^2 D^2 - I_5^{\a\ad,\b\bd}\l(x\r) i \pa_{\b\bd}\bar{D}_\ad D_\a + \l(k-1\r) I_6^{\a\ad}\l(x\r) \bar{D}_\ad D_\a \r] \bar{J} \l(x,\theta,\bar \theta \r) \nonumber \\
\label{totalK} \eea Notice that the sum of the diagrams
$(a),(b),(d),(d^\prime),(e),(e^\prime)$ which contain a vector
propagator reproduces the contributions from the graph $(c)$ which
contains a chiral vertex. Now using the results obtained in
(\ref{oneloop}) we get the ${\cal{O}}\l(g^2\r)$ Konishi-like
correlator \bea
{\cal{K}}_{g^2}\l(0,z\r) & = & \frac{g^2 N}{(2\p)^{2(k+2)}}F_{tree}\frac{k+2}{k} \l[ \l(\bar{D}^2 D^2 + \frac{1}{k+1} i \pa_{a\ad} \bar{D}^\ad D^\a \r) \frac{1}{x^{2\l(k+1\r)}} \, G\l(x\r) \r. \nonumber \\
&& + \bar{D}^2 D^2 \frac{1}{x^{2\l(k+1\r)}} + \frac{3k+1}{2\l(k+1\r)^2} \, i \pa_{\a\ad} \bar{D}^\ad D^\a \frac{1}{x^{2\l(k+1\r)}} \nonumber \\
&& + \l. \frac{1}{2k\l(k+1\r)} \Box \frac{1}{x^{2\l(k+1\r)}} \r]
\d^4\l(\theta\r) \label{finalK} \eea Now we can go back to the
expression in (\ref{formula1loop}) where we have the expected
result for the correlator up to ${\cal{O}}\l(g^2\r)$. A direct
comparison of (\ref{treelevel}) and (\ref{finalK}) with
(\ref{formula1loop}) gives the ${\cal{O}}\l(g^2\r)$ value of the
anomalous dimension \beq \g = \frac{k+2}{k} \ \frac{g^2 N}{4\p^2}
\label{anomdim} \eeq and the normalization of the two-point
function \beq\label{norma} f_{\cal O} =
\frac{1}{{\l(2\p\r)}^{2\l(k+1\r)}} F_{tree} \l[ 1 + \frac{k+2}{k}
\ \frac{g^2 N}{4\p^2} \l( \a + 1 \r) \r] \eeq where $\a$ is the
scheme dependent constant appearing in the function $G\l(x\r)$.
Let's notice that for $k=1$ the value of $\g$ in (\ref{anomdim})
correctly reproduces the known anomalous dimension of the Konishi
operator \cite{AFGJ}. Indeed, for $k=1,2$ our operators
(\ref{konishi}) do not suffer by the mixing problem, so in these
two cases the values in (\ref{anomdim}) are eigenvalues of the
dilation operator (cfr. \cite{B}).

\vspace{0.8cm} We notice that the ${\cal{O}}\l(g^2\r)$ correction to
the normalization of the two-point function is proportional to the
value of the anomalous dimension (this result, while being obvious
for the scheme dependent part in (\ref{norma}), was instead not
expected for the $\a-$independent term in the one-loop
contribution). This suggests that these two objects are related to
each other and, in particular, that protection of the dimension
implies also non-renormalization of the two-point function. This
is in fact the case for the CPO-like operators. The way the final
answer rearranges itself suggests that the result might be
extended to higher orders.

\vspace{1cm}

\noindent {\bf Acknowledgements}

\noindent
We thank M. Boni for help and discussions at an early stage of this work, 
N. Beisert and Y. Stanev for useful comments.

\noindent This work has been partially supported by INFN, MURST, and the European Commission RTN program
HPRN-CT-2000-00113 in which the authors are associated to the University of Torino.

\end{document}